\documentclass[manuscript,screen]{acmart}
\AtBeginDocument{%
  \providecommand\BibTeX{{%
    \normalfont B\kern-0.5em{\scshape i\kern-0.25em b}\kern-0.8em\TeX}}}

\setcopyright{acmlicensed}
\copyrightyear{2024}
\acmYear{2024}
\acmDOI{}

\usepackage{enumitem}
\usepackage{color,soul}

\setlist[itemize]{noitemsep, topsep=0pt}
\begin{document}

\title{Developing Situational Awareness for Joint Action with Autonomous Vehicles}

\author{Robert Kaufman}
\email{rokaufma@ucsd.edu}
\affiliation{%
  \institution{University of California, San Diego}
  \streetaddress{9500 Gilman Dr}
  \city{La Jolla}
  \state{California}
  \country{USA}
  \postcode{92093}
}

\author{David Kirsh}
\email{kirsh@ucsd.edu}
\affiliation{%
  \institution{University of California, San Diego}
  \city{La Jolla}
  \state{California}
  \country{USA}
}

\author{Nadir Weibel}
\email{weibel@ucsd.edu}
\affiliation{%
  \institution{University of California, San Diego}
  \city{La Jolla}
  \state{California}
  \country{USA}
}

\renewcommand{\shortauthors}{Kaufman et al.}

\begin{abstract} %150 words
Unanswered questions about how human-AV interaction designers can support rider's informational needs hinders Autonomous Vehicles (AV) adoption. To achieve joint human-AV action goals - such as safe transportation, trust, or learning from an AV - sufficient situational awareness must be held by the human, AV, and human-AV system collectively. We present a systems-level framework that integrates cognitive theories of joint action and situational awareness as a means to tailor communications that meet the criteria necessary for goal success. This framework is based on four components of the shared situation: AV traits, action goals, subject-specific traits and states, and the situated driving context. AV communications should be tailored to these factors and be sensitive when they change. This framework can be useful for understanding individual, shared, and distributed human-AV situational awareness and designing for future AV communications that meet the informational needs and goals of diverse groups and in diverse driving contexts.
\end{abstract}




%\received{20 February 2024}

\maketitle

\section{Introduction}
Advances in Autonomous Vehicle (AV) technology promise huge individual and societal benefits, from increased driving safety to reduced environmental impact~\cite{fagnant2015preparing}. However, unanswered questions about how AV interaction design can support the critical informational needs of people riding with AVs are hindering real-world adoption~\cite{hegner2019automatic, detjen2021increase} A well documented example is that people often report a lack of trust in AV decision-making, particularly when the AV’s decision procedures are opaque and not human-understandable~\cite{kohn2021measurement, meyer2022baby, morra2019building, ekman2017creating, frison2019ux, wiegand2019drive}.  

One reason we think that the success of AV technologies will likely depend on accurate designing for the social informational needs of people, is because we know that theory of mind is a core part of how we think and move about the world, particularly when we are trying to understand decisions made by other actors~\cite{cuzzolin2020knowing, premack1978does, wang2021towards}. A person seeking to make decisions within a human-AV system necessarily confronts many questions about AV decision-making. Most pressingly for the decisions of the person, questions involve both present and future actions on the road and their predictability: \textit{Why is the AV moving to the left? Does the AV see the pedestrian ahead? What will it do next? Will this always be the case?}

Concern over the effectiveness of human-AI communication may be exacerbated in safety-critical situations~\cite{stanton2001situational, liu2023design} or when a person may have additional interaction goals like learning from the AI’s decisions~\cite{ghai2021explainable, soltani2022user, kaufman2024learning}. AVs of the (not so distant) future may play an expanded role in facilitating more than just safe transportation: managing social, emotional, educational, and cultural experiences for their passengers. For example, an AV could take the scenic route through a city, comment on notable historical centers, and adjust its communication style to avoid waking a sleepy travel companion. AVs could teach driving skills, soothe nervous passengers, and entertain children on a road trip. There are near-endless opportunities. In this paper we provide an illustrative example of what future human-AV interaction may look like, and then outline how it is likely that there will be multiple AI systems working together: some managing the task of driving and analyzing the outside environment, while others will focus on detecting and moderating passenger experiences. Without addressing the salient information needs of passengers, however, very few AV functions can be fully realized. 

\vspace{1em}
\noindent
\textbf{Autonomous Vehicles and Situational Awareness} -- A theory which has helped researchers understand many of the informational problems that stem from interacting with an AV is through the lens of \textbf{situational awareness (SA)}, where people want to be able to perceive, comprehend, and project into the future what is happening in a given driving scenario~\cite{endsley1995toward}. A lack of knowledge in how an AV will behave or why it does so, prevents a person interacting with the AV from achieving situational awareness. This is because they do not have a means to access the information required to develop understanding or predict future behaviors that the AV might take~\cite{capallera2022human}.

\vspace{1em}
\noindent
Lacking situational awareness:
\begin{itemize}
    \item Impedes a person from being able to develop trust, a reliance strategy, or expertise from learning, limiting the usefulness of the technology~\cite{vorm2022integrating, marangunic2015technology}.
    \item Removes a person’s agency, forcing them to rely on blind faith and abandon their desire for understanding~\cite{tulli2024explainable}.
\end{itemize}

\vspace{1em}
\noindent
However, achieving necessary situational awareness in human-AV partnership requires bi-directional communication~\cite{chen2018situation}. Some situational awareness needs to be held by the human and AV independently, while other SA must be shared or distributed between them~\cite{salas1995situation, stanton2006distributed}. Numerous questions remain on how communication needs should be satisfied and how to account for the complex socio-technical system within which the human and AV interact \cite{stanton2006distributed}. 

There have been many models for building situational awareness to support human-AV interaction. These tend to focus on either the needs of the human(s) in the system or the AV in the system independently of one another. Models for the AV’s situational awareness are often fairly technical in nature, aiming to computationally capture what is happening in an external driving situation so that an autonomous vehicle can maneuver safely to a destination~\cite{reich2020sinadra}. Models aimed at building human situational awareness tend to focus on providing justification for an AV's decisions. These range from human-machine interfaces (HMIs) that directly apply Endley’s SA theory for design of in-car communication design~\cite{capallera2022human} to those encompassing aspects of working memory and comprehension~\cite{baumann2007situation, krems2009driving}. There is also a large body of work on how best to measure SA in vehicles~\cite{endsley1988situation, rangesh2018exploring, sirkin2017toward, ma2007situation}. Though useful and informative in their own right, prior work fails to fully account for the joint, system-situated nature of achieving human-AV interaction goals~\cite{kridalukmana2020supportive}. They also fail to show how SA requirements – including what needs to be held individually, shared, or distributed by the humans and AV in the system – may change when aspects of the communicative context changes. 

In this paper we introduce an integrated framework that illustrates how a theory of \textbf{joint action}~\cite{sebanz2006joint} can be used by human-AV interaction designers to understand four foundational factors that affect situational awareness during the moment of human-AV interaction: (1)~joint action goals, (2)~AV traits, (3)~subject-specific traits and states, and (4)~driving context characteristics. We posit that by understanding how these factors form an interdependent system, communications can be designed that meet the informational needs demanded by a particular communicative context. To our knowledge, no prior models or frameworks have examined human-AV communication from the lens of these cognitive and behavioral phenomena. Our framework demonstrates how the perspective of situated human-AV joint action can be used to inform the methods by which we seek to understand SA requirements.

The systems view of human-AV joint action and situational awareness presented here can be used to inform a future research agenda for human-AV interaction research and design. Applying this framework reveals several pressing knowledge gaps which, if filled, can be used to design situationally-aware human-AV systems. We provide specific calls to action for future research. Though framed within the human-autonomous driving domain, many of the knowledge gaps identified likely apply to other human-AI interaction domains, such as those in healthcare or military applications. 

The remainder of this paper is structured as follows. First, we introduce the context of our investigation through an illustrative example (Section ~\ref{example}). Next, we present an overview of our novel framework, which can be used to understand the development of situational awareness supporting situated joint action within Human-AV systems (Section \ref{overview}). In the discussion, we contextualize this framework within existing work and cognitive theory, grounding the framework in examples of driving interaction and providing open areas for future work (Section ~\ref{discussion}); the discussion is broken into three parts: human-AI teaming to achieve action goals (Section ~\ref{goal}), the development situational awareness (Section ~\ref{developing}), and wider system aspects and communication needs (Section ~\ref{wider}). We conclude the paper with an overall description of how this framework can add to the scientific community's understanding of how to design communications meeting situational awareness needs (Section \ref{applying}) and a system summary (Section \ref{conclusion}).

\section{Illustrative Example}
\label{example}

This section contains a fictional story about a small family (mother and son) riding in an AV. Through this example, we illustrate the factors that need to be accounted for when designing communications supporting human-AV interaction goals. This example will be referenced throughout this paper.
\vspace{1em}

\noindent
\textit{A young mother is riding with her 15 year old son in the back seat. They are on vacation and have summoned an AV taxi to help them get from the airport to their hotel. The \textbf{action goals} of the mother are fairly simple: get to the hotel safely. Her son has an additional goal– he is learning to drive, and is interested in understanding how the vehicle maneuvers busy city streets. He loves new technologies and is far more willing to take risks than his mother. Both family members need to rely on the AV to achieve their goals. Trust is a precondition that needs to be addressed to achieve appropriate reliance, which is in itself necessary for safe transportation and learning. The mother and son, their social dynamics and their activities inside the vehicle compose the \textbf{internal driving context}.}

\textit{The \textbf{external driving context} is a traffic-heavy commute on a city street. It’s raining and visibility is low. The vehicle is moving through a busy urban environment, with many intersections and pedestrians. The \textbf{external non-driving context} includes beautiful architecture on either side of the road. \textbf{AV traits} are attributes of the AV which impact the way it acts and can be acted with. The AV’s driving abilities, communication affordances, justification for decisions, and even the algorithms it is trained with are examples of the AV’s traits.}

\textit{In this example, the AV is fully aware of the external driving context via sensors and capable of driving in these conditions safely with the family, so we will assume that the AV is fully capable of performing adequately. For the goal of safe transportation and learning to be satisfied, certain additional actions need to be taken. Specifically, the preconditional criteria to achieve these goals need to be addressed for the human-AV team. These include establishing trust and determining information that can augment the son’s learning experience.}

\textit{Each person in the vehicle has \textbf{subject-specific traits and states} which, for each of their interaction goals, culminates in a set of action needs. The mother is predisposed to distrusting the AV’s decisions: she’s never ridden in an AV before, she is not generally open to new experiences, and she is stressed and fatigued from a long day of travel. On top of this, the presence of her son lowers her tolerance for risk, and the lack of visibility due to rain makes her question the AV's performance in this higher complexity external driving context. Her son is less concerned about the rain than his mom, and is instead intent on understanding why the vehicle is making the decisions that it is: why is it shifting lanes, how does it merge and at what pace does it slow down?}

\textit{Some action needs can be filled without explicit communication: the AV can turn left, slow down, and monitor the external and internal driving contexts. The mother and son can converse, observe the AV's driving, and exit the vehicle.}

\textit{Many action needs are informational in nature and best filled via communication between the people and the AV. The goals, driving context, AV traits, and subject-specific traits and states, and actions combine to form what we call the \textbf{communicative context}. This is a snapshot of the system as a whole, where there are informational needs for all agents that need to be satisfied. If the needs are not satisfied, goals like trust or learning cannot be achieved.}

\textit{Given the case that the AV can actually perform quite well, the AV will need to communicate to the mother sufficient information such that she feels she can trust – and therefore rely on – the AV’s driving decisions. In this way, the AV is building the \textbf{situational awareness} of the mother to include an understanding of its own abilities and actions. To know how to do this, the AV itself needs to either be told by the mother, sense her discomfort and assume it is caused by a lack of trust, or be pre-programmed to communicate information that can increase trust.}

\textit{The AV says via a robotic voice “don’t worry, I am trained to handle this situation and have never crashed.” Though this may be sufficient for the risk-loving 15 year old son, the mother was not fully satisfied by the AV’s first communication– there is still an informational need to fill. She asks, “can you tell me more?”}

\textit{The AV now tries a different \textbf{communicative strategy}. It presents different information via a different modality: on a visual dashboard display, it validates that it can sense all of the vehicles around it. It also shows a projection of where it and other vehicles will drive next so the mother can predict their behavior. This satisfies the mother's need for trust. Now she can form a reliance strategy, thus allowing the goal of safe transportation to be achieved.}

\textit{The son loves the visualization, but seeks further justification for decisions in order to learn. Why is it moving to the left lane? The AV provides further explanations to the son, saying audibly “I am moving to the left lane to prepare for a left turn at the next stoplight.” In this simplistic example, let's say this explanation satisfies the boy’s learning goal.}

\textit{The audible explanations from the vehicle are making the mother's head hurt. She hits a button and the AV switches its instructional explanations to be visual, using text bubbles on the display to continue the son’s instruction. Are these as effective as auditory explanations?}

\textit{A pedestrian crosses in front of the AV and the AV brakes suddenly to avoid collision. The mother’s anxiety spikes: she has new informational needs to fill now.}

\vspace{1em}
\noindent
This example shows how successfully achieving goals and subgoals like trust or learning require coordinated actions between the people and the AV. This, in turn, requires taking into account the information needs of all parties and communicating information in the right way to develop sufficient situational awareness. Only through this situational awareness can informational goals be achieved. When the communicative context changes, so do the SA requirements.

\section{Framework for situated joint action within a human-AV system}
\label{overview}

The framework presented in this paper shows how situational awareness can be developed through joint actions with autonomous vehicles. To summarize: the driving context, human subject traits and states, and AV traits determine actions that need to be taken to achieve specific joint-action goals, such as safe transportation, trust, or learning. Each goal has a success function that determines the conditions that need to be met for that goal to be achieved. The success function is sensitive to the wider context of interaction, meaning when the communicative context changes, so do the conditions for success, so actions must change too. Actions taken to achieve goals and goal criteria may be communicative or non-communicative; some goals have subgoal conditions which need to be met before a higher-level goal can be achieved. Actions are coordinated and bidirectional, meaning that the human and AV need to account for the action needs of each other. To achieve many joint action goals, \textbf{situational awareness (SA)} is necessary. The human and AV both also have \textbf{individual SA}, and when this individual SA matches it is considered to be \textbf{shared SA}. The SA that emerges from the system as a whole is called \textbf{distributed SA}, as the SA possessed by the human and AV does not always need to match, but simply be complimentary with respect to action goals. To achieve a joint action goal, the informational needs required for each type of SA must reach a necessary threshold and be paired with appropriate non-communicative actions. 

\begin{figure}[htp]
  \centering
  \includegraphics[width=\linewidth]{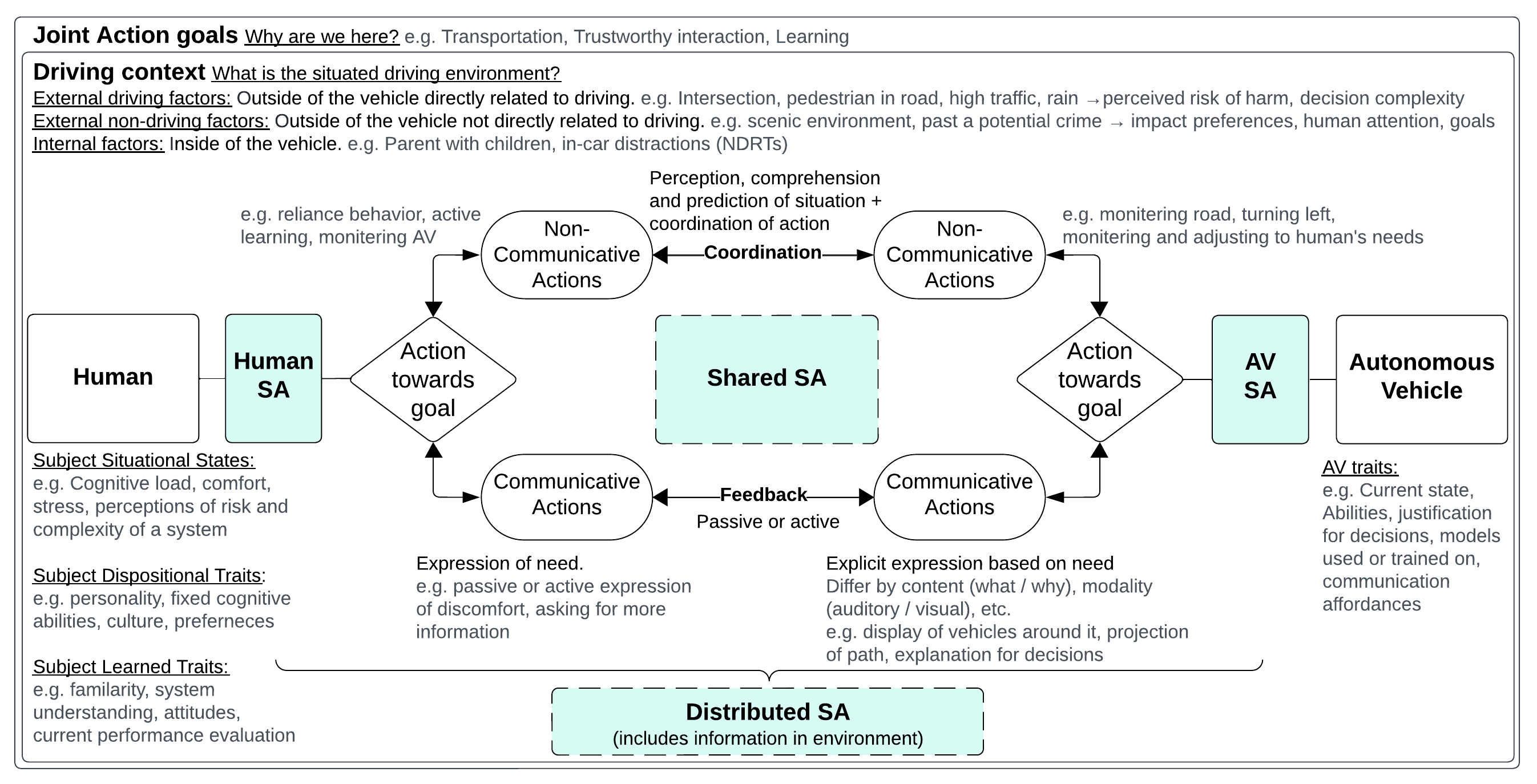}
  \caption{\textbf{Framework overview for situated joint action based on situational awareness within a human-AV system.} Joint Action Goals and AV Traits are discussed in Section ~\ref{goal}, AV traits and the development of situational awareness through strategic communicative actions are discussed in Section ~\ref{developing}, and the driving context and human traits and states are discussed in Section ~\ref{wider}.}
  \label{framework}
\end{figure}

Figure \ref{framework} details our framework, describing the external and internal factors of the human-AV interaction. We show how action goals, the driving context, and different traits (human and AV traits) are fundamental parameters determining what actions - communicative and not - need to be jointly taken. Actions build, and are iteratively impacted by, the different types of situational awareness held by the system. In the next section, we expand on and contextualizes this framework, discussing the specific elements that are at its core.

\section{Discussion}
\label{discussion}
We now discuss the framework in three detailed subsections, contextualizing it in existing work and cognitive theory. The illustrative example presented in Section~\ref{example} will help ground the framework in a realistic driving situation. We first discuss how human-AI teams can work together to achieve joint action goals and subgoals. Then, we show how these goals and subgoals can be achieved through the development of individual, shared, and distributed situational awareness; this development has classically used different communication strategies to communicate various information about the AV (which we term AV traits). In the last part of this section, we discuss how two additional aspects of the human-AV system impact the success criteria (and thus necessary communicative actions) to achieve goals and subgoals. These aspects of the system include the internal and external driving and non-driving contexts as well as the subject-specific traits and states of a human in the loop. Along the way, we provide an overview of related work and knowledge gaps for the human-AV community to address in future work.

\subsection{Human-AI teaming for Action Goal Success}
\label{goal}

As AI-based systems advance, their increasing scope of decision-making actions shifts them from being tools for human use to acting as agents in a human-AI team~\cite{dafoe2021cooperative}. In the case of autonomous vehicles, current theory proposes that successful designs should conceptualize a human and an AV as team members that must work together to achieve common goals, such as safe transportation.

\vspace{1em}
\noindent
\textbf{Action Goals} -- In human-AI teaming, a person and an AV would be bound by principles of joint action similar to those of coordinated human-to-human activities~\cite{sebanz2006joint}. Joint actions are those which involve two or more people or agents working together in coordination to achieve a common goal, and include activities ranging from dressing a child to team rally car driving. For example, the mother and the AV in our example needed to act jointly, in coordination, to achieve the goal of safe transportation. We adopt Sebanz et. al’s definition of \textbf{action goals} as those which direct and orient behavior~\cite{sebanz2006joint}. These include higher-level goals like safe transportation or learning as well as lower-level subgoals like building trust, slowing for a pedestrian, or asking for more information. These subgoals are preconditions that need to be met to achieve higher-level goals; to achieve a higher level goal, the subgoals underneath it need to be addressed. It is clear in our example of the family in the AV: without the coordinated action of both the people and the AV, it would not have been possible to achieve action goals and subgoal criteria.

Some subgoals like building trust in an AV’s decisions have received a lot of focus by the research community. The issue of trust spans nearly all human-AV goals, and is perhaps the most pressing issue hindering AV adoption and the adoption of AI systems more generally~\cite{choi2015investigating}.

Goals like learning from an AV’s decisions are severely underexplored by the research community~\cite{kaufman2024learning}. Knowing how to design for the goal of learning has implications both for the use of AVs as driving instructors as well as the design of communications that can help a person build familiarity, expertise, trust, and comfort with an AV by learning how the AV makes decisions. Learning, like trust, is at the root of many human-AV interaction problems.

\begin{figure}[htp]
  \centering
  \includegraphics[width=0.999\linewidth]{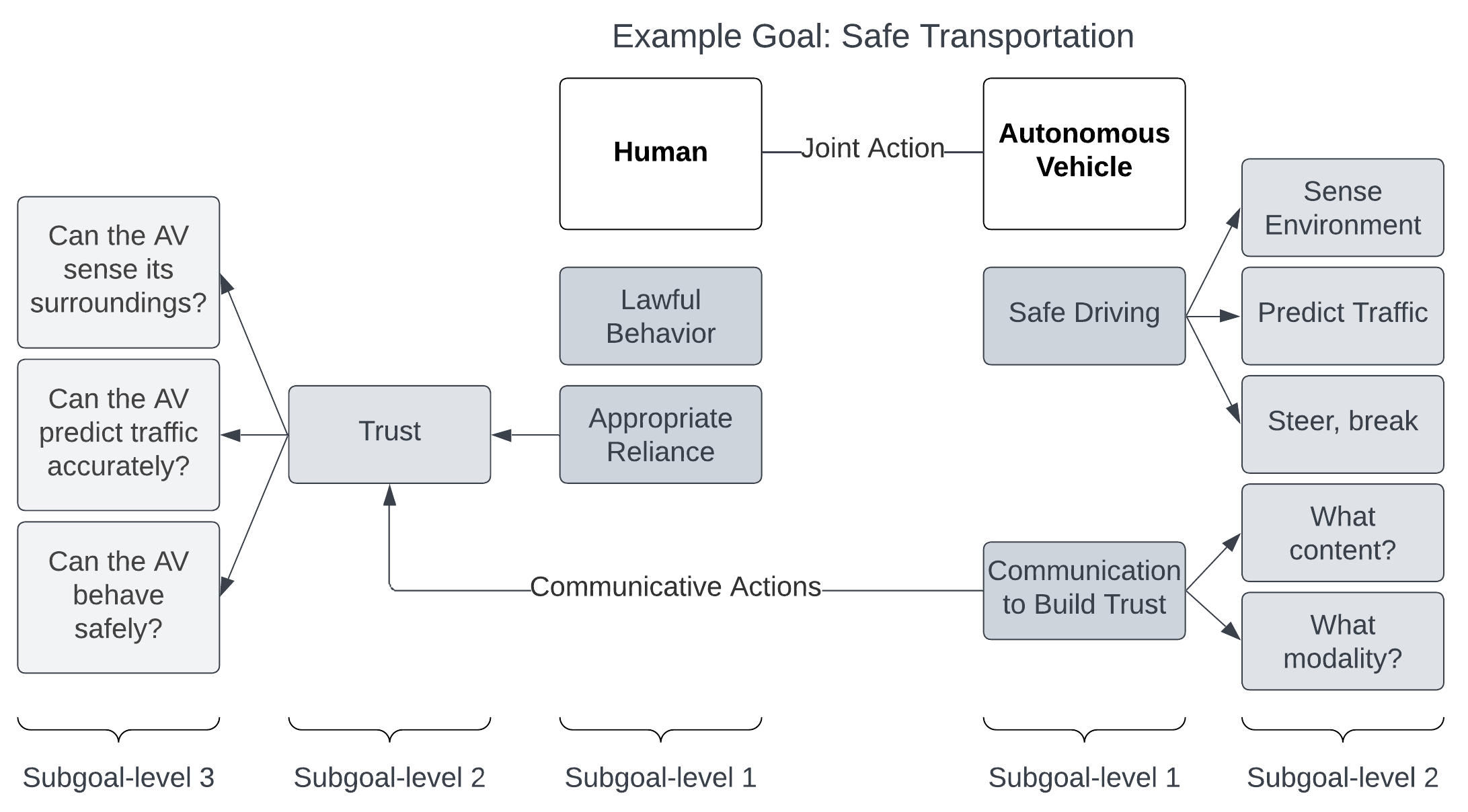}
  \caption{\textbf{Example Success function for the goal of Safe Transportation.} The success function for the goal of safe transportation is composed of sub-goals which need to be met for the overall goal to be achieved. Each of these subgoals have their own conditions, which themselves may have conditions, etc. Joint communicative action may be required between a human and an AV to achieve some goals or subgoals. This example is very simplified and for illustrative purposes only.}
  \label{goals}
\end{figure}

\vspace{1em}
\noindent
\textbf{Success Functions Determine Actions Necessary to Achieve Goals} -- Action goals determine the success function for the system. The success function is composed of a set of criteria which need to be met to achieve that particular goal. We can think of these criteria like a multi-level checklist, where goals have criteria to meet, some of which may be subgoals which, themselves, have their own criteria to meet. There are similar, but not identical to, the goal hierarchies described by Kridalukmana et al. for collaboration with automated vehicles \cite{kridalukmana2020supportive}. Figure \ref{goals} shows an example. To successfully achieve the action goal of safe transportation, for example, the subgoal of calibrated reliance must be met amongst others. To achieve calibrated reliance, knowing when and when not to trust an AV is necessary. To know when and when not to trust an AV, a person needs to be aware of specific information that can be used to develop a trust and reliance strategy. This information has its own criteria, based on principles of communication including the intentional design of the type of information needed to convey as well as how it is conveyed. In many cases, criteria for a particular goal or subgoal cannot be satisfied by a single agent alone; team members must work together to satisfy their function.

For example, the mother in our story wasn’t confident that the AV could perform well in the current rainy driving conditions. She did not trust the AV, and did not know if she could rely on it. To know if she could rely on the AV, she wanted to verify the AV’s capabilities. Specifically, she had criteria of driving competence to fill: \textit{was the AV aware of the other cars around it, and what would it do next?} These are gaps in the mother’s situational awareness. She could not satisfy these criteria alone. To help her address these criteria, the AV produced information on a display which addressed each criteria that could help her build calibrated trust and a reliance strategy. The information produced had their own sub-criteria for success, based on choosing the appropriate content, modality, etc. For the mother and the AV, there was an adjustment period where the first communications produced by the AV did not satisfy the mother’s criteria. This dissatisfaction was communicated back to the AV, letting it know the communications needed to be changed. In the end, because the right information was communicated in the right way, the criteria for success were satisfied, and the joint action goal was achieved. In sum, from a joint-action perspective, the success of a human-AV team working towards a goal depends on members’ ability to (1) share representations, (2) predict the other’s needs and actions, and (3) integrate these predicted actions into their own behavior~\cite{sebanz2006joint}.

\subsection{Developing Situational Awareness}
\label{developing}

\textbf{Individual Situational Awareness and AV Traits} -- The root of satisfying the success criteria for many goals is by satisfying information needs; this is done by developing the \underline{individual situational awareness} of each team member. Chen et al.’s Situation Awareness-Based Agent Transparency (SAT) model of situational awareness for human-agent interaction~\cite{chen2014situation} suggests three categories of agent information are necessary to build a person’s SA: the agent’s current actions and plans, reasoning processes, and outcome predictions. Chen et al.’s updated version of this model (specifically for human-AI teams)~\cite{chen2018situation} adds system transparency and bidirectional communication. Per this updated model, the system’s communications should help a person understand the current state (including execution of actions), reasoning process, and future projections of the system, including how certain the system’s predictions are. We will collectively refer to attributes and affordances of the AV that determine how an AV acts and can be acted with – including these listed by Chen et al. -– as ‘\textbf{AV traits}’. In satisfactorily communicating information related to AV traits, Chen et al. suggest that a person can achieve situational awareness when interacting with AI-based systems. In our family example, the crucial AV traits to be communicated were the driving abilities and decision rationale of the system; communicating these was crucial for helping both the mother and son satisfy their informational goals.

As human cognition and decision-making capacity are necessarily limited, an emphasis on explainability in AI has recently emerged as a key factor supporting successful human-agent interactions. This approach emphasizes that by giving the right information, in the right way, at the right time – an AV can help a person achieve many of their interaction goals, thereby increasing the overall situational awareness that an individual has when driving with their vehicle. This concept is a central fixture for the field of explainable AI (XAI), which emerged in recent years as a means to generate trust with autonomous decision-makers like AVs~\cite{dzindolet2003role, benbasat2005trust}. DARPA’s XAI program posed explainable AI as a means for humans to gain insight into the “black box” nature of machine learned systems so that they know why a decision has been made, when to trust the decision, why errors have occurred, and in what contexts the system will succeed and fail~\cite{gunning2019darpa}. Explainability may be especially important because automation does not have an understanding of cause and effect~\cite{pearl2018book}. Building off this agenda, XAI systems have been used to implement explanations for decisions in many domains~\cite{das2020opportunities, pazzani2022expert, kaufman2022cognitive, kaufman2023explainable}, including autonomous driving~\cite{koo2015did, atakishiyev2021explainable, capallera2022human}.

\vspace{1em}
\noindent
\textbf{Communication Strategies} -- Huge knowledge gaps remain in the human-AV literature on how to provide the right information, in the right way, at the right time to satisfy the criteria for a particular driving goal. For simplicity, we’ll term any communicative action method by which a person and an AV transfer information to eachother during a real-time moment of interaction as ‘\textbf{communication strategies}’. These may include explicit, implicit, or communication via sensing~\cite{capallera2022human}. We are not including data used for prior model training as a communication strategy, though it is relevant to the design of interactive human-AV systems that use this data to decide what communications to produce. To design effective communications, it is important to know both what type of information is most effective for building satisfactory situational awareness as well as what modalities of presentation will most effectively allow access to that information. Choosing the right communication strategy is based on satisfying the information criteria necessary to achieve a joint action goal. To make XAI communication \textit{bi-directional} as Chen et al. suggests, knowledge gaps remain on what information should be provided back to the AV, in what way, and at what time.

In our example, we saw that the first communication strategy (saying “I never crash”) was ineffective at eliciting trust from the mother, while the second communication strategy (display of surroundings and future behavior) was effective for the mother’s trust but ineffective at promoting learning for the son. Different goals and subgoals, people, and contexts will require different communication strategies. This is because different informational success criteria require different informational strategies to meet those criteria. Some studies show visual techniques using heads-up-displays (HUDs) are preferred by people with certain goals in certain driving situations~\cite{schartmuller2019text, chang2016don, kaufman2024learning}, while others suggest auditory feedback as a better strategy~\cite{jeon2009enhanced, locken2017towards}. There have even been a few studies involving haptics specifically for situational awareness during driving~\cite{capallera2019convey, fink2023expanded}. Studies on the impact of the type and content of information presented also show mixed results, including differences between ‘what is happening’ or ‘why it is happening’ information~\cite{koo2015did, omeiza2021not, kaufman2024learning}.

Knowing how to convey information and how information requirements may change for different people or contexts is crucial to designing systems that can meet informational goal criteria. Without careful consideration of these factors, ill-fitting communications may fail to meet rider’s needs or even derail progress towards a goal. We can easily imagine a situation where presenting the wrong information at the wrong time may overwhelm, confuse, or cause a person to lose trust in an AV. These experience would their ability to learn and understand what is happening in a situation.

Until this point, we have discussed situational awareness held by an individual. To understand how to best build human-AV communications that support coordinated joint action, it is helpful to examine situational awareness from two additional perspectives: \textit{shared} situational awareness and \textit{distributed} situational awareness.

\vspace{1em}
\noindent
\textbf{Shared Situational Awareness} -- Shared representation and predictability for joint action (criteria 1 and 2) are quite similar to what would be classically required to achieve \underline{shared situational awareness}. As such, it is possible to examine shared situational awareness within the context of a joint action between a human and an AV. Individual situational awareness becomes shared when there is a \textit{common} representation of a situation where each team member can understand and predict a particular aspect of their environment in synchrony \cite{salas1995situation, nofi2000defining, saner2009measuring}. Individuals with shared situational awareness will have representations of a situation that are similar to each other. For instance, in our story, when the AV displayed the other cars around it, it allowed the family to share a common representation of what the AV perceives with the AV and also with each other. If there was a car in front of the AV that was not displayed, the mother would know that there was a gap in the AV’s situational awareness of the external driving context. This may impact her trust and reliance strategy. Shared situational awareness is often required to satisfy criteria for joint action goals. Shared situational awareness can be at the goal and subgoal establishing level, too. The AV and mother both knowing that there is an informational need to fill to achieve trust is another example of shared situational awareness.

\vspace{1em}
\noindent
\textbf{Distributed Situational Awareness} -- \underline{Distributed situational awareness}, by contrast, proposes that SA is emergent from the system as a whole, rather than held by a given individual \cite{stanton2006distributed, salmon2022distributed, salmon202013}. In terms of a joint action, teams with distributed situational awareness may comprise individuals with different but compatible representations of a situation. By this view, an autonomous vehicle will have its own situational awareness and will also contribute to the total SA of the distributed system as a whole. The distributed view of situational awareness provides a practical way to see how agents with different roles in a joint action team can collaborate on achieving a goal. Using our example, the mother is aware that driving in rain can be risky, that she lacks trust, and when her trust criteria are met. The AV is aware of the distance to other cars on the road, its predictions about their behavior, and how those predictions impact its own decisions. It must also be aware of the informational needs of the mother. In this case, we see that even when internal representations differ – such as within human-AV teams – successful joint action and coordination can still be possible. 

Whether the team is said to have distributed or shared situational awareness may be a matter of shifting perspective between the role of the team and the role of the individual. In most cases, it is likely that both distributed and shared situational awareness may be necessary to achieve a particular action goal. The shift from individual to shared to distributed can best be viewed from the perspective of what is needed to satisfy action goal criteria. For example, both a human and an AV may need to share an accurate representation of the “big picture” for a particular driving context – including an understanding of the joint subgoal criteria of trust– while the information required to \textit{achieve} this subgoal can be distributed between independent members. To achieve trust, specific decisions regarding whether to trust the AV may be delegated to the mother, while accurately calculating the distance to the nearest car is left to the AV \cite{salmon202013}. Of course, these are interrelated, as communications given to the mother by the AV about the nearest car will affect whether her criteria for trust are satisfied. The implication is that designing for human-AV situational awareness is possible despite fundamental differences in the modalities of representation employed by humans and AI-based systems like AVs, provided appropriate information is communicated in accessible ways to each. The awareness held by each, and the communicative actions taken by each, are relative to achieving the action goal.

Crucial to applying this work for human-AV interaction designers, it is important to once again emphasize that joint actions are coordinated and bidirectional: each partner must integrate their partner’s needs and behaviors in space and time to their own action planning (joint action criteria 3). Action coordination affects the affordances of what can be done, motivates what should be done, and permits joint anticipatory adjustment of future behavior to achieve a joint goal. This means that to achieve trust, the mother, AV and system as a whole must have sufficient situational awareness – and ability to act on this awareness – such that they can account for the needs and actions of the other.

\subsection{Wider System Factors Impacting Success Criteria and Communication Needs}
\label{wider}

\vspace{1em}
\noindent
\textbf{Situated Driving and Non-Driving Contexts} -- The human-AV system does not exist in a vacuum; it is sensitive to the situated, contextual demands of the physical and social environment~\cite{clancey1997situated, kirsh2009problem}. We use ‘\textbf{driving context}’ as an umbrella term encompassing the scene, situation, and scenario within which the human and AV are interacting~\cite{ulbrich2015defining}. Succinctly, a scene is a momentary snapshot of elements like traffic and an agent’s self-representation of these elements; a situation includes all relevant conditions, options and determinants for action derived from a scene via information selection and goal-driven processes; and a scenario is a sequence of scenes. The driving context includes both factors outside of the vehicle and inside the vehicle.

The \textbf{external driving context} includes features like road conditions, traffic, and visibility~\cite{capallera2019owner}. These may impact the perceived risk of harm or cognitive complexity of knowing what to do in a particular situation. In our example, the external driving context  was a high traffic commute, in an urban city, during rain. This can be further divided by the specific commute features, like moving through an intersection or stopping for an unexpected pedestrian. 

The \textbf{external non-driving context} includes factors outside of the vehicle which do not necessarily have direct implications on driving, but may impact the driving experience itself. In our example, this includes the sight of beautiful building architecture outside of the vehicle, which does not impact the AV’s driving but may impact passenger preferences and how attention is directed throughout the ride. Predicting how the external non-driving context will impact passenger states and goals requires general intelligence. 

The \textbf{internal driving context} may include social dynamics between riders or distractions that may arise inside the vehicle cabin, such as Non-Driving Related Tasks (NDRTs)~\cite{naujoks2018review}. We adopt a situated action view, where the driving context includes viewing driving as a social, culturally-dependent activity where cognitive processes are distributed across the system~\cite{hollan2000distributed}. The driving context from our example changed when the unexpected pedestrian appeared in the road, requiring the AV to stop. It would also change if another person entered the vehicle (internal) or if the rain stopped (external).

As the driving context changes, so do the criteria required to satisfy the success function for a particular goal. It is clear that joint actions between a human and an AV must take many aspects of the distributed system into account in order to allow the team to reach their goals. Several studies have incorporated external driving conditions into their situational awareness studies~\cite{capallera2022human, de2020designing}, however nearly all focus on identification of obstacles via interfaces~\cite{sirkin2017toward} with very little attention given to internal driving contexts or adapting communication between different driving contexts. Underlying aspects of the driving contexts -– like risk of harm and situation complexity -– which may help meaningfully categorize contexts to give a deeper insight into the situated context, are likewise unexplored. Open research questions remain regarding how contextual changes impact the informational criteria of particular goals or subgoals like learning or developing trust in an AV. \textit{What contextual factors matter, to what degree, and how they should be communicated about?} Understanding the potentially differential impact of context on communication needs is paramount to building context-aware systems that can work even when the driving situation changes~\cite{schilit1994context}.

\vspace{1em}
\noindent
\textbf{Subject-specific Traits and States} -- To coordinate joint actions, an AV will need a representation of the needs, behaviors, and attributes of their human team member(s), just as the person needs a representation of the AV’s traits. We call the attributes that determine needs and behaviors for a particular person their ‘\textbf{subject-specific traits and states}’, which will vary from person to person. In building trust with automated systems, Hoff and Bashir’s model divides subject-specific factors into three categories: situational, dispositional, and learned~\cite{hoff2015trust}. Though these are not specifically for the development of situational awareness, they are useful as a scaffold for understanding the wide range of subject-specific factors which may need to be accounted for in designing situationally-aware and adaptable human-AV systems. According to Hoff and Bashir, situational states vary by a specific situation and include a person’s perceptions of a system’s risk and complexity, cognitive load, confidence, and subject-matter expertise. The latter three – cognitive load, confidence, and expertise – are where a majority of human-AV situational awareness studies have focused on~\cite{capallera2022human, saner2009measuring}. Subject-specific traits that are dispositional (generally fixed) in nature, including measures of culture, personality, gender, cognitive style, and information processing abilities are fairly underexplored~\cite{ferronato2020examination}. The same can be said for learned factors stemming from prior experiences and preexisting knowledge. Learned factors include technology attitudes, expectations, system understanding, and changes based on a momentary evaluation of a system’s performance \cite{ayoub2021modeling}.

It is not difficult to see how variance in subject-specific traits and states may impact the informational needs people may develop. For example, the mother in our story had never ridden in an AV before, was not generally open to new experiences, and was stressed and fatigued from a long day of travel. Her risk-willingness was also lower due to the presence of her child. We can imagine that she may have different informational needs than a single, open-minded tech lover who enjoys risk-taking. We would also expect that there may be differences in her needs between her first time interacting with an AV and her hundredth. 

Open questions remain regarding which subject-specific traits and states may be relevant, and to what degree they need to be represented. A sufficient understanding of the impact of subject-specific traits can be used to design flexible, tailored communications for diverse audiences. Integrating these together during model training can be used to customize communications to a specific person.

\vspace{1em}
\noindent
\textbf{Communications Should Be Sensitive To The Communicative Context} -- The AV’s communications should be sensitive to the human they are interacting with as well as the relevant goals and contextual factors, causing a complex interaction between actions (including communication strategies), AV traits, subject-specific traits and states, action goals, and the driving context. For simplicity, we collectively refer to these factors in combination as the ‘\textbf{communicative context}’. We can think of a single communicative context as a momentary snapshot of the full driving system, which determines the criteria which need to be satisfied for a particular goal as well as the ways in which that goal can be satisfied. When an aspect of the system changes – such as when a new passenger gets into the AV or the vehicle enters into a new driving context -– the communicative context changes. In some cases, the criteria for success will change as well as the possible ways to satisfy that criteria.

With the right AV communications, an AV may be able to deliver information such that \textit{sufficient} situational awareness is achieved, and joint actions taken by the team can reach a team’s action goals. What counts as \textit{sufficient} is based on the success function for the specific action goal in question – informational criteria for a goal may be met even with partial situational awareness. For different communicative contexts, the proportion of necessary individual, shared, and distributed situational awareness may change. Success criteria for particular action goals are not often clearly articulated by the research community. However, many preliminary XAI studies on human-AV interaction show promise that certain AV representations can help increase human trust, understanding, and situational awareness. Open questions remain regarding how much situational awareness is necessary for each team member to possess to permit joint action and goal success. In addition, the degree to which different factors (like subject-specific traits or communication strategies) may impact the amount of situational awareness obtained remains unknown. For example, it is possible that people who are predisposed to trusting automation will require less situational awareness when interacting with a vehicle. It is also possible that XAI communications will be more effective for open-minded novices or for those with low cognitive load. There is likely a nonlinear relationship between the amount of information and the usefulness of that information, dependent on information content. In essence, there are dozens of potential avenues for future work.

\section{Applying the Framework}
\label{applying}
This framework can be used to show how accounting for AV traits, action goals, subject-specific traits and states, and the driving context can be used to design communications that can satisfy human-AV situational awareness success criteria for joint action goals like learning from an AV or conditional subgoals like building trust. In an applied sense, breaking the human-AV system into its constituent parts (e.g. subject-specific traits, contextual characteristics) and emphasizing interactions between these parts (e.g. trust in an AV may be lower in rain) provides a scaffold that can be used by future researchers and designers to understand many complex human-AI interaction problems in a diverse range of domains. For real-world deployment of autonomous vehicles and most other AI systems, ensuring that interactions meet the needs of their (often diverse) user bases and can remain contextually-relevant when goals or contexts change can be a distinct design challenge. This framework can be used as a tool to find combinations of factors that need to be accounted for when designing these interactions. Past and future research can determine what factors are important to consider, and when. If we design a system without taking each aspect into account, AI communications run the risk of failing to meet the needs users. This may alienate a user or - in the case of AVs and other safety-critical systems - cause real-world physical or financial harm.

One procedure that interaction designers can take is to work through the framework systematically. Starting with the goals of the human-AV team, an initial success function can be written (similar to the example in Figure \ref{goals}, but more comprehensive) showing what actions are needed to achieve that goal and its subgoals. What is known about the AV traits, subject traits and states, and driving context can be included from the beginning as a means to determine the human-AV team's needs, with knowledge gaps highlighted as areas needed to be filled. For example, a designer might know the information content desired by a particular user for a particular goal and in a particular driving context, but may not know the \textit{amount}, \textit{modality}, or \textit{timing} of the information to be presented, or how information should be adjusted should a new driving context or user enter the system. As information is collected about what is known and not, the subgoals and subgoal criteria can be iteratively adjusted in the success function in order to show what actions should be taken. It is unrealistic to assume that all aspects of the system can be accounted for at all times, so an emphasis should be on what is \textit{necessary} and \textit{sufficient} to include. Determining the importance of different factors and what should be included will be part of the design challenge. Different functions may need to be drawn for different user types, contexts, or goals. Through this exercise, designers can gain a comprehensive mapping for how to design communications that meet a success function's criteria and/or an understanding of what additional research is needed.

In line with the process used for designing interactions outlined above, this framework can also serve as a foundation to deepen our study of situational awareness and joint action within complex human-AI systems. A similar process of walking through success criteria and determining what is known, unknown, and important from each of the system factors can be used by the research community to articulate gaps in knowledge that must be filled before AI systems can reach their full potential (e.g. need to know how to design communications for low-trust people in the rain). From the standpoint of cognitive theory, experiments targeting specific areas of the system can help differentiate what type and amount of SA - individual, shared, or distributed - is necessary for different conditions and what the most efficient methods are for developing this SA within different communicative contexts. Showing how SA requirements may differ for people with high or low cognitive load capacity, for example, can give additional insight into the relationship between cognitive load and working memory. 

\section{Conclusion: Framework and System Summary}
\label{conclusion}
In this work, we present a framework for developing situational awareness within human-autonomous vehicle interactive systems based on a theory of joint action. To summarize the framework: individual, shared and distributed situational awareness – up to a particular threshold – are necessary for a human-AV team to satisfy the success criteria to jointly achieve action goals. Joint action goals include high-level goals such as safe transportation and learning from an AV’s behavior, which themselves often have lower-level subgoals like trust which need to be met. In many cases, both the human and the AV must have a sufficiently accurate shared representation of each other and the situated driving context. They also must have the individual situational awareness required to coordinate and complete each of their own distinct actions. The situational awareness of the distributed system as a whole must be sufficient that the team together can accomplish action goals. Information can be transferred via communication strategies like XAI explanations for a person, and real-time human input and sensing for an AV. Training data can be used by an AV as a means to predict and tailor communications for a specific communicative context ahead of time. Understanding four intersecting components of the shared situation – AV traits, action goals, subject-specific traits and states, and the driving context – is necessary to promote situational awareness and permit joint action. Successful joint action requires the extra step of predicting and adjusting behavior based on individual and shared situational awareness to reach the team’s goal. For human-AV XAI design, the implication is that AV communications should be tailored based on these four factors and be sensitive when the communicative context changes. Huge knowledge gaps remain in each of these areas.

\begin{acks}
We would like to thank Emi Lee, Janzen Molina, Saumitra Sapre, and Rohan Bhide for their support on projects related to human-AV interaction which inspired the work in this manuscript. We would like to thank Cat Hicks for feedback related to theoretical components of this framework.
\end{acks}

\bibliographystyle{ACM-Reference-Format}
\bibliography{bibliography}

\end{document}